\def\dfrac#1#2{{\displaystyle\frac{#1}{#2}}}

\def\beq{\begin{equation}}
\def\eeq{\end{equation}}
\def\bea{\begin{eqnarray}}
\def\eea{\end{eqnarray}}

\def\non{\nonumber}

\documentclass[a4j,12pt]{article}
\usepackage[dvips]{graphicx}

\begin{document}

\pagestyle{empty}

\thispagestyle{empty}


\begin{center}
{\Large {\bf Determination of $S_{17}$ from systematic

\vspace{-1mm}

analyses on $^8$B Coulomb breakup with

\vspace{1.5mm}

the Eikonal-CDCC method
}}

\vspace{3mm}

Kazuyuki Ogata, M. Yahiro$^{\rm A}$, Y. Iseri$^{\rm B}$,
T. Matsumoto, N. Yamashita and\\ M. Kamimura

\vspace{3mm}

{\it Department of Physics, Kyushu University

$^{\rm A}$Department of Physics and Earth Sciences, University of the Ryukyus

$^{\rm B}$Department of Physics, Chiba-Keizai College
}

\end{center}

\vspace{3mm}

\begin{abstract}
A new version of the method of Continuum-Discretized Coupled-Channels
method (CDCC) is proposed, that is, the Eikonal-CDCC method (E-CDCC).
The E-CDCC equation, for Coulomb dissociation in particular,
can easily and safely be
solved, since it is a first-order differential equation and contains
no huge angular momentum in contrast to the CDCC one.
The scattering amplitude calculated by E-CDCC has a similar form to
that by CDCC. Then one can construct hybrid amplitude in an intuitive way,
i.e., CDCC amplitude is adopted for a smaller angular momentum $L$
and E-CDCC one for a larger $L$ related to an impact parameter $b$.
The hybrid calculation is found to perfectly reproduce the quantum
mechanical result
for $^{58}$Ni($^8$B,$^7$Be$+p$)$^{58}$Ni at 240 MeV, which shows
its applicability to systematic analysis of $^8$B dissociation to
extract the astrophysical factor $S_{17}$ with high accuracy.
\end{abstract}

\section{Introduction}
\label{introduction}

The solar neutrino problem is one of the central issues in
the neutrino physics~\cite{Bahcall}.
Nowadays, the neutrino oscillation is assumed
to be the solution of the problem and the solar neutrino
physics moves its focus on determining oscillation parameters:
the mass difference among $\nu_{e}$, $\nu_{\mu}$ and $\nu_{\tau}$,
and their mixing angles~\cite{Bahcall2}.
The astrophysical factor $S_{17}$,
defined by $S_{17}(E)\equiv \sigma_{p\gamma}(E)E\exp [2\pi \eta]$ with
$\sigma_{p\gamma}$ the cross section of the $p$-capture reaction
$^7$Be($p,\gamma$)$^8$B and $\eta$ the Sommerfeld parameter,
plays an essential role in the parameter-search procedure,
since the prediction value for the flux of $^8$B neutrino, which
is intensively being detected on the earth, is proportional to $S_{17}(0)$.
The required accuracy from astrophysics is about 5\% in errors.

Because of difficulties of direct measurements for the $p$-capture reaction
at very low energies, alternative indirect measurements
were proposed. $p$-transfer
reaction~\cite{Azhari,Trache,Ogata}
with the Asymptotic Normalization Coefficient (ANC) method~\cite{Xu}
and $^8$B Coulomb dissociation~\cite{RIKEN,GSI,MSU}
are typical examples of them;
we concentrate on the latter in the present paper.
$^8$B Coulomb dissociation can be assumed to be the inverse
reaction of $^7$Be($p,\gamma$)$^8$B provided the $^8$B is dissociated
through only its E1 transition by absorption of virtual photons.
Once this condition is satisfied, cross section of the $p$-capture
reaction, thus $S_{17}(E)$, can easily be obtained by applying the
principle of detailed balance to the measured dissociation cross sections.
Intensive measurements of the $^8$B Coulomb dissociation are being made
in RIKEN~\cite{RIKEN}, GSI~\cite{GSI} and MSU~\cite{MSU};
the extracted $S_{17}(0)$ are almost consistent
each others. There is, however, contradiction between MSU and RIKEN
data about the contribution of the E2 component. Additionally,
role of nuclear interaction, interference between nuclear
and Coulomb interactions in particular, has not yet clarified quantitatively.
Therefore, it can be said that the extracted $S_{17}(0)$ from those $^8$B
Coulomb dissociation measurements contain uncertainties to some extent.

In order to determine $S_{17}(0)$ accurately, careful analysis
of the $^8$B dissociation including both nuclear and Coulomb interactions
is necessary. The method of Continuum-Discretized Coupled-Channels
(CDCC)~\cite{CDCC},
which was proposed and developed by Kyushu group,
is suitable for that purpose. In fact, it was shown in Ref.~\cite{MSU}
that CDCC can very well
reproduce the MSU data. It is not straightforward, however, to extract
the component corresponding to the inverse reaction of $^7$Be($p,\gamma$)$^8$B
out of the total measured spectra.

Very recently~\cite{Yamashita},
we proposed a procedure that determines $S_{17}(0)$ from
the $^8$B dissociation measurements by using the ANC method,
which is free from uncertainties by the use of the detailed balance.
An important advantage of the CDCC + ANC analysis is that one
can quantitatively evaluate the accuracy of the extracted $S_{17}(0)$;
the fluctuation of ANC by changing the $^8$B single particle
wave functions is interpreted as the error of $S_{17}(0)$ coming from
the use of the ANC method.
We analyzed
$^{58}$Ni($^8$B,$^7$Be$+p$)$^{58}$Ni at 25.8 MeV measured at
Notre Dame~\cite{ND},
for which detailed balance was found to fail to obtain $S_{17}(0)$
because of its low incident energy~\cite{EB}.
The extracted $S_{17}(0)$ is
22.83 $\pm$ 0.51 (theo) $\pm$ 2.28 (expt) eVb and the ANC method
turned out to work very well, i.e., less than 1\% of error; the remaining
theoretical error comes from the choice of the modelspace of CDCC calculation.
Although quite large systematic error (10\%) of the experimental
data prevents one from determining $S_{17}(0)$ with the required accuracy,
this method significantly reduces theoretical errors of $S_{17}(0)$.

The CDCC + ANC analysis is expected to be applicable to
the RIKEN and MSU data at several tens of MeV/nucleon.
From practical point of view, however, CDCC calculation
including long-ranged Coulomb coupling-potentials requires
extremely large modelspace rather difficult to handle;
typically the required number of partial waves is 15,000 for
the MSU data~\cite{MSU}.
Although interpolation technique for angular momentum reduces the
number of Coupled-Channels (CC) equations to be solved,
CC equation with huge angular momentum $L$ is rather
unstable and careful treatment is necessary.
In this sense, it seems almost impossible to apply CDCC to the GSI data
at 250 MeV/nucleon, where more than 100,000 partial waves
will be required.

In the present paper we propose a new version of CDCC; the Eikonal-CDCC
method (E-CDCC) being based on eikonal and non-adiabatic
calculation.
Since E-CDCC equations contain no angular momentum for the relative motion
of a projectile and a target nucleus, the $S$ matrix is obtained
easily and safely.
Another important feature of the method is that
the resultant scattering amplitude has the form which is very similar
to that obtained by CDCC, by using the relation between
impact parameter $b$ and $L$.
This allows one
{\it hybrid} calculation, i.e., one can adopt the amplitude of E-CDCC
only for larger $L$, where classical picture is expected to be valid,
and {\it connect} them with that of the standard,
fully quantum mechanical, CDCC for smaller $L$.
The hybrid calculation includes all quantum-mechanical effects necessary
through the CDCC amplitudes.
In addition to that, the use of the E-CDCC amplitudes
removes all problems concerning with huge angular momenta and
drastically reduces computation time.
From theoretical point of view,
the hybrid calculation will give an insight into the {\it connection}
between quantum-mechanical and classical pictures.

In Sec.~\ref{formalism} formalism of E-CDCC and the construction of
hybrid scattering amplitude are briefly described.
We show in Sec.~\ref{result} the validity of the hybrid calculation
for $^{58}$Ni($^8$B,$^7$Be$+p$)$^{58}$Ni at 240 MeV; results for
the elastic and total breakup cross sections are compared with
those of fully quantum-mechanical calculation. Effects of
adiabatic approximation and importance of the interference between
the larger- and smaller-$L$ regions are also discussed.
Finally summary and conclusions are given in Sec.~\ref{summary}.

\section{Formalism}
\label{formalism}

In this section we briefly describe formalism of E-CDCC and how
to construct hybrid scattering amplitude by using a result of
(standard) CDCC partly.
Detailed formalism and theoretical foundation of CDCC are shown
elsewhere~\cite{CDCC,Piya,CDCC-foundation}.

%
%
\begin{figure}[htbp]
\begin{center}
\includegraphics[width=70mm,keepaspectratio]{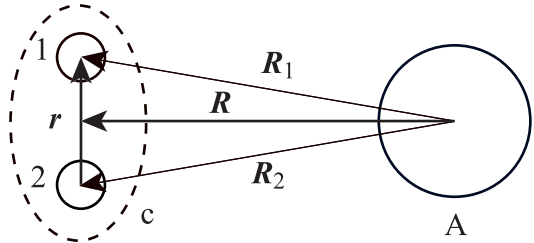}
\end{center}
\vspace*{-5mm}
\caption{
Schematic illustration of the system concerned in the present paper.
}
\end{figure}
The system that we treat in the present study is illustrated in Fig.~1.
We start with the expansion of the total wave function $\Psi$:
\beq
\Psi({\bf R},{\bf r})
=
\sum_{i \ell m}
\Phi_{i,\ell m}({\bf r})
\sum_{m_0}
e^{-i(m-m_0)\phi_R}
\chi_{i \ell m}(R,\theta_R),
\label{psinew2}
\eeq
where $\ell$ is the total spin of c and $m$ is its projection
on the $z$-axis taken to be parallel to the incident beam; the
subscript 0 represents the initial state.
$\Phi_{i,\ell m}$ is the channel wave function of c, which
contains both bound and scattering states and
for the latter discretization of the continuum is made.
\{$\Phi_{i,\ell m}$\} is assumed to form an approximate complete set
for a finite configuration space being significant for a reaction
concerned.

We here make the following eikonal approximation:
\beq
\chi_{c}(R,\theta_R)
\approx
\psi_{c}(b,z)
\dfrac{1}{(2\pi)^{3/2}}
e^{i{\bf K}_c(b) \cdot {\bf R}},
\label{eikonal}
\eeq
where $c$ denotes channels \{$i$, $\ell$, $m$\} together and
the wave number $K_c$ is defined by
\beq
\dfrac{\hbar^2}{2\mu}\left(K_c(b)\right)^2=
E-\epsilon_{i,\ell}-\dfrac{\hbar^2}{2\mu}
\dfrac{(m-m_0)^2}{b^2}
\label{capk}
\eeq
with $b$ the impact parameter;
the direction of ${\bf K}_c$ is assumed to be parallel to the $z$-axis.

Inserting Eqs.~(\ref{psinew2}) and (\ref{eikonal}) into the three-body
Schr\"{o}dinger equation neglecting
the second order derivative of $\psi_{c}$, one can obtain
\beq
\dfrac{i\hbar^2}{\mu}
K_c^{(b)}
\dfrac{d}{d z}\psi_{c}^{(b)}(z)
=
\sum_{c'}
{\cal F}^{(b)}_{cc'}(z)
\;
\psi_{c'}^{(b)}(z)
e^{i\left(K_{c'}^{(b)}-K_c^{(b)}\right) z},
\label{cceq4}
\eeq
where ${\cal F}^{(b)}_{cc'}$ is the form factor, i.e.,
the interaction between A and each constituent of c, folded by
$\Phi_{c}$ and $\Phi_{c'}$;
we put $b$ in a superscript since
it is not a dynamical variable but an input parameter.
Equation (\ref{cceq4}) is solved with the boundary condition
$\psi_{c}^{(b)}(-\infty)=\delta_{c0}$.
Since the E-CDCC equation is first-order differential one and
contains no coefficient with huge angular momentum, one can easily
and safely solve the equation.

Using the solution of the E-CDCC equation, the
scattering amplitude with E-CDCC is given by
\beq
f_{c0}^{\rm E}
=
-\dfrac{\mu}{2\pi\hbar^2}
\int
\sum_{c'}
{\cal F}^{(b)}_{cc'}(z)
\,
e^{-i(m-m_0)\phi_R}
e^{i ({\bf K_{c'}^{(b)}}-{\bf K_c'^{(b)}})\cdot {\bf R}}
\,
\psi_{c'}^{(b)}(z)
d{\bf R},
\label{f}
\eeq
where $\mu$ is the reduced mass of the c + A system.
Making use of the following forward-scattering approximation:
\bea
e^{i ({\bf K_{c'}^{(b)}}-{\bf K_c'^{(b)}})\cdot {\bf R}}
&=&
-K_c^{(b)}\sin{\theta_f}b\cos{\phi_R}+(K_{c'}^{(b)}-K_c^{(b)}\cos{\theta_f})z
\non \\
&\approx&
-K_c^{(b)}\theta_f b\cos{\phi_R}+(K_{c'}^{(b)}-K_c^{(b)})z,
\label{C}
\eea
one obtains
\beq
f_{i \ell m,i_0 \ell_0 m_0}^{\rm E}
=
\dfrac{1}{2\pi i}
\int
\!\!
\int
K_{i \ell m}^{(b)}
e^{-i(m-m_0)\phi_R}
e^{-iK_{i \ell m}^{(b)} \theta_f b\cos{\phi_R}}
\left(
{\cal S}_{i \ell m,i_0 \ell_0 m_0}^{(b)}
\!\!-
\delta_{i i_0}\delta_{\ell \ell_0}\delta_{m m_0}
\right)
bdbd\phi_R,\\
\label{f4}
\eeq
where
the eikonal $S$-matrix elements are defined by
${\cal S}_{i \ell m,i_0 \ell_0 m_0}^{(b)}\equiv
\psi_{i \ell m}^{(b)}(\infty)$.

We here {\it discretize} $f^{\rm E}$:
\bea
f_{i \ell m,i_0 \ell_0 m_0}^{\rm E}
&=&
\dfrac{1}{2\pi i}
\sum_L
K_{i \ell m}^{(b_L^{\rm mid})}
\left[
\int
e^{-i(m-m_0)\phi_R}
e^{-iK_{i \ell m}^{(b_L^{\rm mid})}\theta_f b_L\cos{\phi_R}}
d\phi_R
\right]
\non \\
& &
\times{}
\left(
{\cal S}_{i \ell m,i_0 \ell_0 m_0}^{(b_L^{\rm mid})}
-
\delta_{i i_0}\delta_{\ell \ell_0}\delta_{m m_0}
\right)
\int_{b_L^{\rm min}}^{b_L^{\rm max}}
bdb,
\label{f6}
\eea
where
$b_L^{\rm min}$, $b_L^{\rm max}$ and $b_L^{\rm mid}$
are defined through
$K_{i \ell m}^{(b_L^{\rm min})}b_L^{\rm min}= L$,
$K_{i \ell m}^{(b_L^{\rm max})}b_L^{\rm max}= L+1$
and $K_{i \ell m}^{(b_L^{\rm mid})}b_L^{\rm mid}= L+1/2$,
respectively. In deriving Eq.~(\ref{f6}) we
neglecting the $b$-dependence of
$K_{i \ell m}^{(b)}$,
$\exp[-iK_{i \ell m}^{(b)}\theta_f b\cos{\phi_R}]$
and
${\cal S}_{i \ell m,i_0 \ell_0 m_0}^{(b)}$
within small size bin of $b$ corresponding to each $L$.
After manipulation one can obtain
\beq
f_{i \ell m,i_0 \ell_0 m_0}^{\rm E}
\!\!\approx
\dfrac{2\pi}{i K_0}
\sum_L
\dfrac{K_0}{K_{i \ell m}^{(b_L^{\rm mid})}}
\sqrt{\dfrac{2L+1}{4\pi}}
i^{(m-m_0)}
Y_{L \,m\!-m_0}(\hat{\bf K}')
\left(
{\cal S}_{i \ell m,i_0 \ell_0 m_0}^{(b_L^{\rm mid})}
\!\!-
\delta_{i i_0}\delta_{\ell \ell_0}\delta_{m m_0}
\right),
\non \\
\label{f7}
\eeq
which has a similar form to that of standard CDCC:
\bea
f^{\rm Q}_{i \ell m,i_0 \ell_0 m_0}
\!\!\!\!\!&=&\!\!\!
\dfrac{2\pi}{iK_0}
\sum_{L}
\sum_{J=|L-\ell|}^{L-\ell}
\sum_{L_0=|J-\ell_0|}^{J-\ell_0}
\sqrt{\dfrac{2L_0+1}{4\pi}}
(L_0 0 \ell_0 m_0 | J m_0)
(L \;m_0\!\!-\!m \;\ell m | J m_0)
\non \\
\!\!\!& &\!\!\!
\hspace{16mm}
{}\times
(S_{i L \ell,i_0 L_0 \ell_0}^{J}-\delta_{i i_0}
\delta_{L L_0}\delta_{\ell \ell_0})
(-)^{m-m_0}Y_{L \,m\!-m_0}(\hat{\bf K}').
\label{fq}
\eea

The construction of the hybrid scattering amplitude $f^{\rm H}$
is rather straightforward:
\beq
f_{i \ell m,i_0 \ell_0 m_0}^{\rm H}
\equiv
\sum_{L\le L_{\rm C}}
f_L^{\rm Q}
+
\sum_{L > L_{\rm C}}
f_L^{\rm E},
\label{fh}
\eeq
where $f_L^{\rm Q}$ ($f_L^{\rm E}$) is the $L$-component of
$f^{\rm Q}$ ($f^{\rm E}$).
$L_{\rm C}$ represents the connecting point between
quantum-mechanical and classical pictures, which is chosen
so that $f_L^{\rm E}$ coincides with $f_L^{\rm Q}$ for
$L>L_{\rm C}$.
It should be noted that Eq.~(\ref{fh}) includes all quantum-mechanical
effects being necessary and also interference between the two $L$-regions.

In the above formulation we neglect the Coulomb distortion.
In order to include it, we use
$
\chi_{c}^{J}(R,\theta_R)
\approx
\chi'^{J}_{c}(b,z)
\phi_{c}^{\rm C}(b,z)
\label{eikonal2}
$
instead of Eq.~(\ref{eikonal}), where $\phi_{c}^{\rm C}$ is
the Coulomb wave function~\cite{Kawai}.
The formulation of $f^{\rm E}$ can then be done just in the same way
above.

\section{Results and discussion}
\label{result}

In order to see the validity of the hybrid calculation with
CDCC and E-CDCC, we analyze $^{58}$Ni($^8$B,$^7$Be$+p$)$^{58}$Ni
at 240 MeV.
Parameters of the modelspace are as follows.
The numbers of bin-states of $^8$B
are 16, 8 and 8 for s-, p- and d-states, respectively.
The maximum excitation energy of $^8$B is 10 MeV and
$r_{\rm max}$ ($R_{\rm max}$) is 100 fm (500 fm).
$L_{\rm max}$ is taken to be 1000 for saving computation time,
which is somewhat small compared with
the expected value of $L_{\rm max}$ that gives perfect convergence,
i.e.,  $L_{\rm max}\sim 4000$.
%
%
\begin{figure}[bh]
\begin{center}
\includegraphics[width=140mm,keepaspectratio]{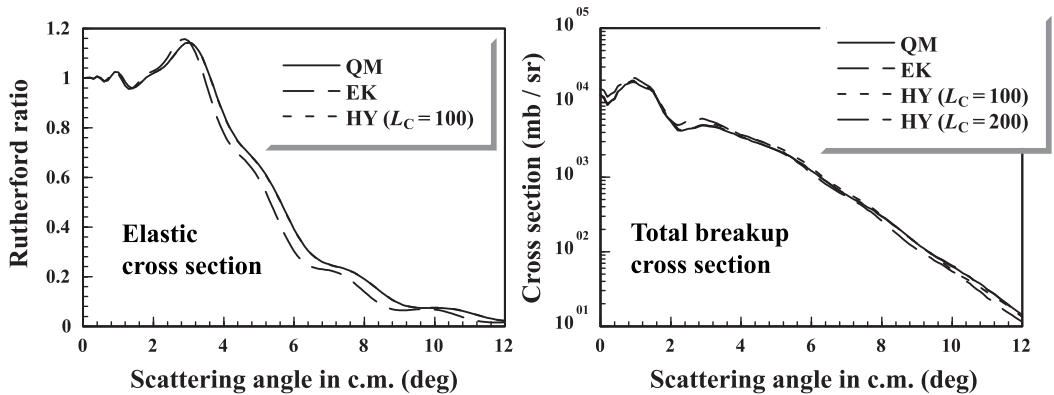}
\end{center}
\vspace*{-5mm}
\caption{
Angular distribution of
the elastic (left panel) and total breakup (right panel)
cross sections for
$^{58}$Ni($^8$B,$^7$Be$+p$)$^{58}$Ni at 240 MeV.
The solid, dashed and dotted lines
show, respectively, the results of quantum-mechanical (QM),
eikonal (EK) and hybrid (HY) calculation with $L_{\rm C}=100$.
The dash-dotted line in the right panel
represents HY calcualtion with $L_{\rm C}=200$.
}
\end{figure}
However, the validity of the hybrid calculation can definitely
be clarified in the modelspace above.
As for $^8$B wave functions, the single-particle model by
Kim {\it et al.}~\cite{Kim} was adopted.
For nuclear interaction between $^7$Be ($p$) and $^{58}$Ni
we used the parameter set of Cook et al.~\cite{Cook}
(Becchetti and Greenlees~\cite{BG});
we neglected the spin-dependent part for simplicity.
Additionally, we replaced Eq.~(\ref{capk}) by
$(\hbar^2/2\mu)K_c^2=E-\epsilon_{i,\ell}$ and neglected $c$-dependence
of $K$ in the calculation of $f^{\rm E}$, which was found to have
no effects on numerical results.
We stress that this is not an adiabatic approximation since $K_c$ is
explicitly treated in solving the E-CDCC equation (\ref{cceq4}).

In the left and right panels in Fig.~2 we show
the elastic cross section (Rutherford ratio) and total breakup cross
section as a function of scattering angle in center-of mass (c.m.)
frame, respectively.
The solid, dashed and dotted lines
represent the results of quantum-mechanical (QM),
eikonal (EK) and hybrid (HY) calculation, where the $L_{\rm C}$ are
taken to be 1000, 0 and 100.
In the right panel the result of HY calculation with $L_{\rm C}=200$
is also shown by the dash-dotted line.
The agreement between QM and HY calculation with appropriate
value of $L_{\rm C}$, namely 100 (200) for elastic (breakup) cross
section, is excellent;
the error is only less than 1\%.
One also sees fairy large difference between EK and QM calculation.
Since our EK calculation contains no corrections to the straight-line
approximation,
this does not directly show the fail of EK approximation.
However, it seems quite difficult for EK calculation to
obtain {\lq\lq}perfect'' agreement with the result of QM one.
On the contrary, the HY calculation turned out to be applicable to analyses
of $^8$B dissociation where very high accuracy is required.

%
%
\begin{figure}[ht]
\begin{center}
\begin{minipage}[t]{.45\textwidth}
\includegraphics[width=65mm,keepaspectratio]{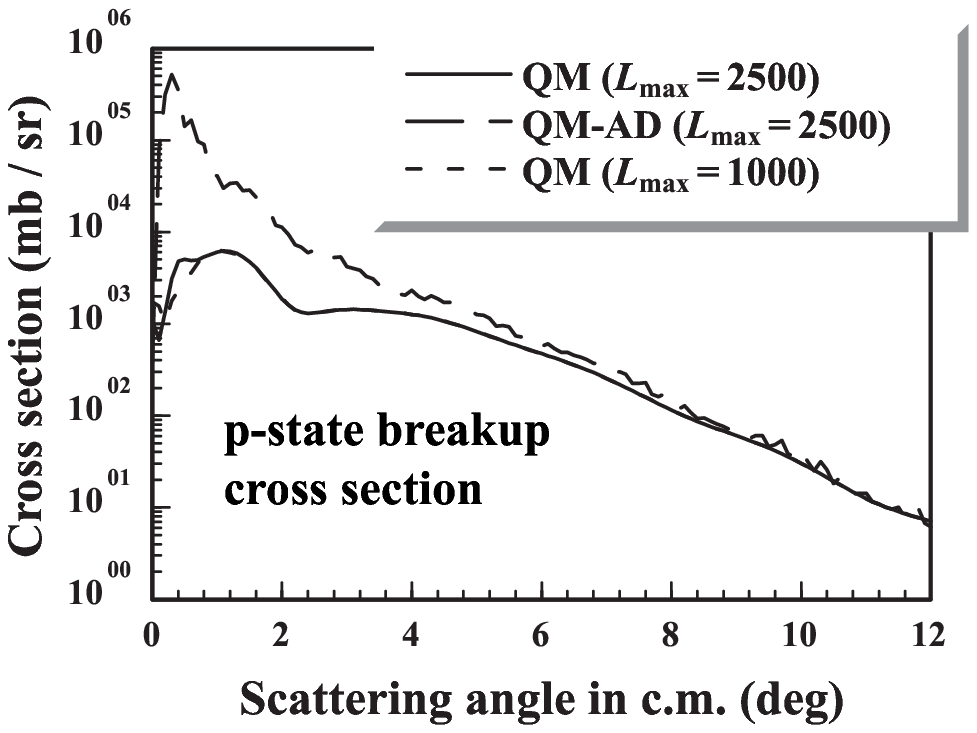}
\caption{
The p-states breakup cross sections by QM calculation
with (dashed line)
and without (solid line) adiabatic (AD) approximation.
The maximum number of $L$ is taken to be 2500.
For comparison non-adiabatic QM result with $L_{\rm max}=1000$
is shown by the dotted line.
}
\end{minipage}
\begin{minipage}[t]{.2\textwidth}
\end{minipage}
\begin{minipage}[t]{.45\textwidth}
\includegraphics[width=70mm,keepaspectratio]{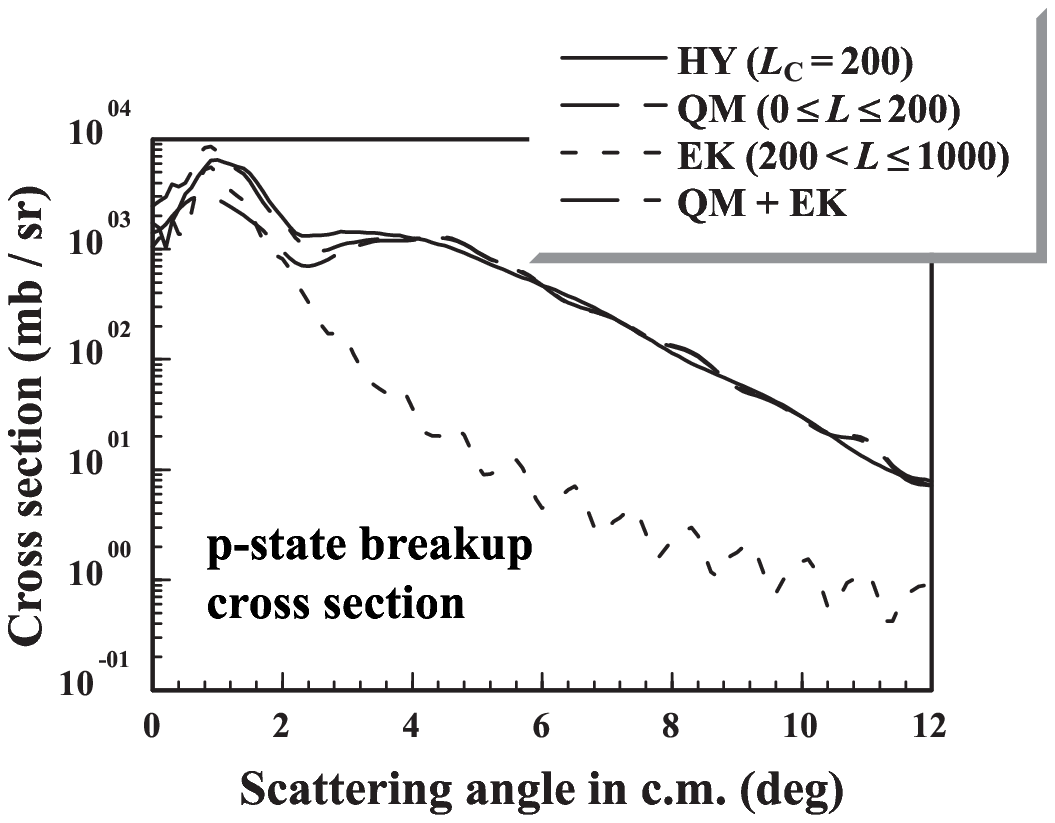}
\caption{
The p-state breakup cross sections by
QM calculation with $0 \le L \le 200$ (dashed line)
and by EK calculation with $200< L \le 1000$ (dotted line).
The dash-dotted line is the incoherent sum of the
dashed and dotted lines and the solid line is the
HY result with $L_{\rm C}=200$, namely, the coherent sum of the two.
}
\end{minipage}
\end{center}
\end{figure}
We show in Fig.~3 the p-state breakup cross sections by
QM calculation with (dashed line) and without (solid line)
adiabatic (AD) approximation.
Since the convergence of AD calculation is rather slow
compared with non-AD one, we take $L_{\rm max}=2500$
to make fair comparison.
One sees that AD calculation much overestimates the breakup cross
section, which shows the failure of AD approximation for this reaction.
In Fig.~3 the non-AD QM result with $L_{\rm max}=1000$ is also shown by
the dotted line. The difference between the solid and dotted
lines is finite but quite small, which indicates the QM calculation
with $L_{\rm max}=2500$ works well, i.e.,
the large difference between AD and non-AD results is indeed
due to the use of AD approximation.

In Fig.~4 the p-state breakup cross sections by
QM calculation with $0 \le L \le 200$ and EK calculation with
$200< L \le 1000$ are shown by the dashed and dotted lines,
respectively.
The dash-dotted line is the incoherent sum of the two,
which deviates from the HY result
shown by the solid line. This result shows that the essence of
our HY calculation is the construction of the hybrid scattering
amplitude not the hybrid cross section.

\section{Summary and Conclusions}
\label{summary}

In the present paper we propose a new method to treat breakup process
accurately and efficiently, that is, the hybrid (HY) calculation with
the Continuum-Discretized Coupled-Channels method (CDCC) and
the Eikonal-CDCC method (E-CDCC). E-CDCC describes the
center-of-mass (c.m.) motion between the projectile and the target nucleus
by straight-line (eikonal approximation) and treats
the excitation of the
projectile explicitly, by constructing discretized-continuum states
as in CDCC, i.e., non-adiabatic and non-perturbative calculation
can be done. When Coulomb distortion appears, we make a similar
approximation to the eikonal one; we use Coulomb wave functions
instead of plane waves.
The resultant scattering amplitude by E-CDCC has a similar form
to the quantum-mechanical (QM) one obtained by CDCC, which
allows one to construct HY amplitude $f^{\rm H}$
in an intuitive way;
$f^{\rm H}$ is given by the sum of the partial amplitude
calculated by CDCC for smaller angular momentum $L$ and that by
E-CDCC for larger $L$, i.e., larger impact parameter $b$.

We analyzed $^{58}$Ni($^8$B,$^7$Be$+p$)$^{58}$Ni at 240 MeV by
QM, full-eikonal (EK) and HY calculation. It was found that
the elastic and total breakup cross sections obtained by QM
calculation are {\lq\lq}perfectly'' reproduced by the HY calculation,
namely, the error is only less than 1\%.
Constructing HY cross section, not HY amplitude, turned out
to fail to reproduce the corresponding QM result, which shows
the importance of the interference between the two $L$-regions above.

In conclusion, the hybrid calculation using CDCC and E-CDCC
allows one very accurate analyses for breakup processes.
The accuracy of the model is enough to be applied to $^8$B
dissociation relating to the astrophysical factor $S_{17}(0)$,
the aim of which is determining $S_{17}(0)$ with less than
5\% errors. E-CDCC drastically reduces computaion time
and eliminates many problems concerned with huge
angular momentum in solving Coupled-Channels (CC) equations.
Thus, the hybrid calculation opened the door to the systematic analyses
of $^8$B dissociation measured at RIKEN, MSU and GSI.
Extracted $S_{17}(0)$ by using the ANC method will be reported
in near future.

\section*{Acknowledgement}

The authors wish to thank M. Kawai, T. Motobayashi and T. Kajino
for fruitful discussions and encouragement.
We are indebted to the aid of JAERI and RCNP, Osaka University
for computation.
This work has been supported in part by the Grants-in-Aid for
Scientific Research
of the Ministry of Education, Science, Sports, and Culture of Japan
(Grant Nos.~14540271 and 12047233).


\begin{thebibliography}{00}

\bibitem{Bahcall}
J. N. Bahcall {\it et al.},
Astrophys. J. {\bf 555}, 990 (2001) and references therein.

\bibitem{Bahcall2}
J. N. Bahcall {\it et al.},
JHEP {\bf 0108}, 014 (2001)
[arXiv:hep-ph/0106258];
JHEP {\bf 0302}, 009 (2003)
[arXiv:hep-ph/0212147].

\bibitem{Azhari}
A. Azhari {et al.},
Phys. Rev. C {\bf 60}, 055803 (1999);
Phys. Rev. Lett. {\bf 82}, 3960 (1999).

\bibitem{Trache}
L. Trache {et al.},
Phys. Rev. Lett. {\bf 87}, 271102 (2001).

\bibitem{Ogata}
K. Ogata {\it et al.},
Phys. Rev. C {\bf 67}, R011602 (2003).

\bibitem{Xu}
H. M. Xu {\it et al.},
Phys. Rev. Lett. {\bf 73}, 2027 (1994).

\bibitem{RIKEN}
T. Motobayashi {\it et al.},
Phys. Rev. Lett. {\bf 73}, 2680 (1994);
T. Kikuchi {\it et al.},
Eur. Phys. J. A {\bf 3}, 209 (1998).

\bibitem{GSI}
N. Iwasa {\it et al.},
Phys. Rev. Lett. {\bf 83}, 2910 (1999).

\bibitem{MSU}
B. Davids {\it et al.},
Phys. Rev. Lett. {\bf 86}, 2750 (2001);
Phys. Rev. C {\bf 63}, 065806 (2001).

\bibitem{CDCC}
M. Kamimura {\it et al.},
Prog. Theor. Phys. Suppl. {\bf 89} (1986);
N. Austern {\it et al.},
Phys. Rep. {\bf 154}, 125 (1987).

\bibitem{Yamashita}
N. Yamashita, Master thesis, Kyushu University, 2003.

\bibitem{ND}
J. von Schwarzenberg {\it et al.},
Phys. Rev. C {\bf 53}, 2598 (1996);
J. Kolata {\it et al.},
Phys. Rev. C {\bf 63}, 024616 (2001).

\bibitem{EB}
H. Esbensen and G. F. Bertsch,
Phys. Rev. C {\bf 59}, 3240 (1999).

\bibitem{Piya}
R. A. D. Piyadasa {\it et al.},
Phys. Rev. C {\bf 60}, 044611 (1999).

\bibitem{CDCC-foundation}
N. Austern {\it et al.},
Phys. Rev. Lett. {\bf 63}, 2649(1989);
N. Austern {\it et al.},
Phys. Rev. C {\bf 53}, 314 (1996).

\bibitem{Kawai}
M. Kawai, private communication (2003).

\bibitem{Kim}
K. H. Kim {\it et al.},
Phys. Rev. C {\bf 35}, 363 (1987).

\bibitem{Cook}
J. Cook,
Nucl. Phys. {\bf A388}, 153 (1982).

\bibitem{BG}
F. D. Becchetti and G. W. Greenlees,
Phys. Rev. {\bf 182}, 1190 (1969).

\end{thebibliography}
\end{document}